\documentclass[prd,aps,nofootinbib]{revtex4}

\def \BEA { \begin{eqnarray}}
\def \EEA {\end{eqnarray}}
\def \BE {\begin{equation}}
\def \EE {\end{equation}}

\def \WDS #1 {\mbox{$\Phi_{#1}^{S}$}}
\def \WDA #1 {\mbox{$\Phi_{#1}^{A}$}}
\def \WD #1 {\mbox{$\Phi_{#1}$}}

\newcommand{\kphi}[2] {\phi_{#1}^{(#2)}}
\newcommand{\kpi}[2] {\pi_{#1}^{(#2)}}
\newcommand{\kC}[2] {c_{#1}^{(#2)}}
\newcommand{\kpsi}[2] {\psi_{#1}^{(#2)}}

\newcommand{\ta}[2] {\tau_{#1}^{(#2)}}

\def \mi {\stackrel{i}{m}}
\def \mj {\stackrel{j}{m}}
\def \mk {\stackrel{k}{m}}
\def \mr {\stackrel{r}{m}}
\def \ms {\stackrel{s}{m}}
\def \mp {\stackrel{p}{m}}
\def \mz {\stackrel{z}{m}}
\def \mq {\stackrel{q}{m}}
\def \mo {\stackrel{o}{m}}
\def \mD {\stackrel{2}{m}}
\def \mT {\stackrel{3}{m}}
\def \mC {\stackrel{4}{m}}

\def \mio #1 {\mi_{#1}\ ^{  \! \! \! \! 0}} 
\def \mjo #1 {\mj_{#1}\ ^{  \! \! \! \! 0}} 
\def \mko #1 {\mk_{#1}\ ^{  \! \! \! \! 0}} 
\def \mro #1 {\mr_{#1}\ ^{  \! \! \! \! 0}} 
\def \mso #1 {\ms_{#1}\ ^{  \! \! \! \! 0}} 
\def \mpo #1 {\mp_{#1}\ ^{  \! \! \! \! 0}} 
\def \mzo #1 {\mz_{#1}\ ^{  \! \! \! \! 0}} 
\def \mqo #1 {\mq_{#1}\ ^{  \! \! \! \! 0}} 
\def \moo #1 {\mo_{#1}\ ^{  \! \! \! \! 0}} 
\def \mDo #1 {\mD_{#1}\ ^{  \! \! \! \! 0}} 
\def \mTo #1 {\mT_{#1}\ ^{  \! \! \! \! 0}} 
\def \mCo #1 {\mC_{#1}\ ^{  \! \! \! \! 0}} 

\def \miJ #1 {\mi_{#1}\ ^{  \! \! \! \! (1)}} 
\def \mjJ #1 {\mj_{#1}\ ^{  \! \! \! \! (1)}} 
\def \mkJ #1 {\mk_{#1}\ ^{  \! \! \! \! (1)}} 
\def \mrJ #1 {\mr_{#1}\ ^{  \! \! \! \! (1)}}

\def \bl {\mbox{\boldmath{$\ell$}}}

\def \bn {\mbox{\boldmath{$n$}}}

\def \hbm #1 {\mbox{\boldmath{$\hat m^{(#1)}$}}}
\def \bm {\mbox{\boldmath{$m$}}}

\def \Mi {\stackrel{i}{M}}

\def \Ms {\stackrel{s}{M}}

\newcommand{\be}{\begin{equation}}
\newcommand{\ee}{\end{equation}}
\newcommand{\beqn}{\begin{eqnarray}}
\newcommand{\eeqn}{\end{eqnarray}}
\newcommand{\pa}{\partial}
\newcommand{\ba}{\begin{array}}
\newcommand{\ea}{\end{array}}

\def \BEAH {\begin{eqnarray*}}
\def \EEAH {\end{eqnarray*}}
\def \BEA {\begin{eqnarray}}
\def \EEA {\end{eqnarray}}
\def \BDM {\begin{displaymath}}
\def \EDM {\end{displaymath}}
\def \Mi {\stackrel{i}{M}}

\def \Ms {\stackrel{s}{M}}

\begin{document}

\title{On asymptotically flat algebraically special spacetimes in higher dimensions}

\author{M. Ortaggio}
\email{ortaggio(at)math(dot)cas(dot)cz}
\author{V. Pravda}%
 \email{pravda@math.cas.cz}
 \author{A. Pravdov\'a}%
 \email{pravdova@math.cas.cz}
\affiliation{
Institute of Mathematics, Academy of Sciences of the Czech Republic \\ \v Zitn\' a 25, 115 67 Prague 1, Czech Republic}

\date{\today} 

\begin{abstract}
We analyze asymptotic properties of higher-dimensional vacuum spacetimes admitting a ``non-degenerate'' geodetic multiple WAND. After imposing a fall-off condition necessary for asymptotic flatness, we determine the behaviour of the Weyl tensor as null infinity is approached along the WAND. This demonstrates that these spacetimes do not ``peel-off'' and do not contain gravitational radiation (in contrast to their four-dimensional counterparts). 
In the non-twisting case, the uniqueness of the Schwarzschild-Tangherlini metric is also proven.
\end{abstract}

\pacs{04.50.-h, 04.20.Ha, 04.20.-q}

\maketitle

\section{Introduction}

Asymptotically flat spacetimes describe the gravitational field of isolated systems in general relativity, and their behaviour near (spacelike or null) infinity encodes essential information about physical quantities such as mass, angular momentum, and flux of radiation. The development of relevant techniques has proven fundamental in understanding general properties of the theory, since it enables one to characterize large classes of solutions in a unified way. However, in recent years the interest in higher dimensional gravity has grown considerably (see, e.g., \cite{EmpRea08} and references therein). This was motivated by modern unified theories, AdS/CFT and recent brane world scenarios, which opened a possible direct link to rich and qualitatively new observable phenomenology. Notions such as the total energy of an isolated system and energy flux are thus fundamental also in higher dimensional theories \cite{HolIsh05,HolWal04,Ishibashi08}. In addition, since properties of gravitational waves depend on the model under consideration (in particular, on spacetime dimensions \cite{CarDiaLem03}), the study of radiation in higher dimensions may ultimately enable one to distinguish different models. 

It was noticed a long ago \cite{Sachs61} that four-dimensional algebraically special spacetimes, while leading to significant mathematical simplification, still asymptotically retain the essential features of (outgoing) radiation fields generated by more realistic sources. As discussed above, it is now natural to explore similar ideas for $n>4$ spacetime dimensions. The recently developed $n>4$ generalization of the Petrov classification \cite{Coleyetal04}  provides us with the necessary formalism. According to the possible existence of WANDs (Weyl aligned null directions) 
and their multiplicity, spacetimes of arbitrary dimension are  invariantly classified into the principal types 
 G, I, II (D), III, N, and O (type II specializes to the secondary type D when {\em two} distinct multiple WANDs are present). {As in four dimensions,} more  special types lead in general to simpler field equations, when expressed in the generalized Newman-Penrose formalism \cite{Pravdaetal04, OrtPraPra07}. 
Hence, it is the purpose of the present paper to analyze higher-dimensional asymptotically flat vacuum spacetimes of type II (or more special) as a first step towards a more general study.
This choice is also motivated by the fact that while in four dimensions a number of explicit such solutions is known (e.g., the Kerr metric and its accelerated generalizations, radiative Robinson-Trautman solutions, etc. \cite{Stephanibook}), a corresponding catalog is at present much poorer for $n>4$. {Our conclusions will also partly explain why it is so, at least in the non-twisting case.}

Apart from algebraical types, one can also  characterize algebraically special spacetimes in terms of geometrical properties of the null congruence(s) generated by the WAND(s).
Under quite general conditions a {\em multiple} WAND of a vacuum spacetime must be geodetic \cite{PraPraOrt07,Pravdaetal04}. Moreover, it has been proven recently that any $n>4$ Einstein spacetime that admits a non-geodetic multiple WAND must also admit a geodetic one \cite{DurRea09}, and we shall restrict our analysis to the case of {\em geodetic} multiple WANDs. Geometrical properties of {geodetic} null congruences are encoded in the $(n-2) \times (n-2)$ matrix $L$ and, in particular, in its invariants expansion, shear, and twist \cite{Pravdaetal04, OrtPraPra07} (see the definitions below). {Since we are interested in asymptotically flat spacetimes,} in the following we will assume $\det L\neq 0$, i.e. that the matrix $L$ is ``non-degenerate''.\footnote{A rigorous proof that WANDs of asymptotically flat spacetimes must satisfy such a condition needs to be given. This seems to be, however, quite plausible, since $\det L=0$ means that some spacelike dimensions are geometrically preferred, which is rather typical, e.g., of Kaluza-Klein asymptotics (for instance, the field generated by black strings/black branes, see \cite{OrtPraPra09}).} 
{In view of the results of \cite{Pravdaetal04}, we can already anticipate that this condition will rule out spacetimes of type III and N, leaving us with the genuine type II (D) only. This does not appear to be a major limitation here,} 
since type II (D) is the simplest principal class allowing for black holes \cite{PraPraOrt07}, and indeed the well known Myers-Perry black holes \cite{MyePer86} belong to this class \cite{Hamamotoetal06,PraPraOrt07}. 

For clarity, let us summarize our assumptions: We study asymptotically flat vacuum spacetimes admitting a geodetic multiple WAND $\bl$ with $\det L\neq0$. We follow the notation of \cite{Coleyetal04,Pravdaetal04,OrtPraPra07,PraPraOrt07}, and employ a frame consisting of two null vectors $\bm_{(0)}=\bl$ 
 (aligned with the multiple WAND) and 
 $\bm_{(1)}=\bn$, 
 and $n-2$ orthonormal spacelike vectors $\bm_{(i)}$, 
 with frame indices $0,1$ and $i, j, \dots=2,\ldots,n-1$. Derivatives along the frame vectors $\bl$, $\bn$ and $\bm_{(i)}$ are denoted by $D$, $\Delta$ and $\delta_i$, respectively.
The frame will be {\em parallely transported} along $\bl$. The optical matrix $L$ of $\bl$ has matrix elements $L_{ij}=\ell_{a;b}m_{(i)}^am_{(j)}^b$, with (anti-)symmetric parts $S_{ij}=L_{(ij)}$ and $A_{ij}=L_{[ij]}$, and other Ricci rotation coefficients are defined similarly \cite{Pravdaetal04,OrtPraPra07}. The optical scalars expansion, $\theta $, shear, $\sigma$, and twist, $\omega$, are defined by $\theta =L_{ii}/(n-2)$,
$\sigma^2=(S_{ij}-\theta \delta_{ij})(S_{ij}-\theta \delta_{ij})$, $\omega^2=A_{ij}A_{ij}$.
For the frame components of the Weyl tensor we use the compact symbols 
\be
 \WD{ij} = C_{0i1j} , \quad \Psi_{i} = C_{101i}, \quad \Psi_{ijk}= \frac{1}{2} C_{1kij}, \quad \Psi_{ij} = \frac{1}{2} C_{1i1j} ,
\ee
denoting by $\WDS{ij} $, $\WDA{ij} $, and $\WD{ } \equiv \WD{ii} $ the symmetric and antisymmetric parts
of $\WD{ij} $ and its trace, respectively. From the symmetries and the tracelessness of the Weyl tensor one has the identities $
C_{01ij}=2 C_{0[i|1|j]}=2\WDA{ij} $, $2C_{0(i|1|j)}=2\WDS{ij} =-C_{ikjk}$, $2C_{0101}= -C_{ijij}=2\WD{ } $, $\Psi_i=2 \Psi_{ijj}$, $\Psi_{\{ijk\}}=0$, $\Psi_{ijk}=-\Psi_{jik}$, $\Psi_{ij}=\Psi_{ji}$, and $\Psi_{ii}=0$, which will be employed throughout the paper.

\section{General asymptotics}

By exploiting the interplay between the Ricci and Bianchi equations \cite{Pravdaetal04,OrtPraPra07}, which govern the ``propagation'' of the Ricci rotation coefficients and of the Riemann (Weyl) tensor, we determine in this section the dependence of the Weyl tensor on an affine parameter $r$ along the geodetic multiple WAND $\bl=\pa/\pa r$. The results will be discussed in the subsequent final section.

\subsection{Sachs equation}

The starting point of our analysis is the higher-dimensional Sachs equation {for $L$, since this matrix will then enter all the relevant Bianchi equations. Under the above assumptions} it reads (see eq.~(11g) of \cite{OrtPraPra07})
\BE
 DL=-L^2 .
 \label{matrixSachs}
\EE

Since $\det L\neq 0$, eq.~(\ref{matrixSachs}) can be rewritten as $L^{-1}(DL)L^{-1}=-I$, where $I$ is the identity matrix. Then using $0=D(L^{-1}L)=(DL^{-1})L+L^{-1}DL$ we immediately get $DL^{-1}=I$, so that (cf.~\cite{NP})
\be
 L^{-1}=r{I}-b , 
 \label{inverse}
\ee
where the ``constant'' matrix $b$ (i.e., $Db=0$) specifies the ``initial conditions''.
By taking the inverse matrix, eq.~(\ref{inverse}) uniquely fixes the $r$-dependence of $L=(rI-b)^{-1}$. In particular, since $L^{-1}_{[ij]}=-b_{[ij]}$, we have 
\be
 A_{ij}=0 \Leftrightarrow b_{[ij]}=0 ,
 \label{twistfree}
\ee
i.e. $b_{[ij]}$ is responsible for the twist of $\bl$ {(but it also enters $S_{ij}$)}.

From eq.~(\ref{inverse}), the behaviour of $L$ for large $r$ follows (after restoring matrix indices)
\be
 L_{ij}=\sum_{m=0}^p\frac{(b^m)_{ij}}{r^{m+1}}+{\cal O}(r^{-p-2})=\frac{1}{r}\delta_{ij}+\frac{1}{r^2}b_{ij}+{\cal O}(r^{-3}) , 
 \label{L_infty}
\ee
which, to the leading order, simply becomes $L_{ij}\approx r^{-1}\delta_{ij}$ {(and the expansion $\theta$ is clearly non-zero)}.

\subsection{Asymptotics of the Weyl tensor}

We can now fix the $r$-dependence of the Weyl tensor by integrating the Bianchi identities containing $D$-derivatives. As it turns out, it is convenient to consider them in order of decreasing boost weight, i.e., in the sequence $0$, $-1$, $-2$ (components of positive boost weight vanish by our assumptions). 

\subsubsection{Components of boost weight zero}
\label{bw_zero}

Since $\bl$ is { geodetic} the relevant Bianchi equations for zero boost weight components decouple from those for negative boost weight and can thus be solved independently. In particular eqs.~(B5) and (B12) of \cite{Pravdaetal04} reduce to
\BEA
 D \WD{ij} &=& -  \WD{} L_{ij} -\WD{ik} L_{kj}
- 2\WDA{ik}  L_{kj} , \label{DPhiij} \\
 DC_{ijkm}&=& -\WD{kj} L_{im}+\WD{mj} L_{ik}-\WD{mi} L_{jk} +\WD{ki} L_{jm}-C_{ijkl}L_{lm}+C_{ijml}L_{lk}-4\WDA{ij} A_{km}\label{DCijkm} .
\EEA

In order to study the asymptotics ($r\to\infty$) of these components, we expand them in non-positive powers of $r$ (positive powers would lead to a p.p.~curvature singularity and can thus be excluded). Then we have 
\BE
 \WD{ij} = \sum_{m=0}^p (\ta{ij}{m} +\kpi{ij}{m}) r^{-m} + {\cal O}(r^{-(p+1)}) , \qquad    
 C_{ijkl} = \sum_{m=0}^p \kC{ijkl}{m}  r^{-m} + {\cal O}(r^{-(p+1)}) , 
 \label{Phi_symm}
\EE
where, from now on, symbols with a superscript $^{(m)}$  do not depend on $r$,
$\ta{ij}{m}=\ta{ji}{m}$, $\kpi{ij}{m}= -\kpi{ji}{m}$, 
$\kphi{}{m}\equiv\ta{ii}{m}$, 
and $2\ta{ij}{m}=-\kC{ikjk}{m}$. 
Substituting these expressions into the Bianchi equations (\ref{DPhiij}) (multiplied by $L^{-1}_{jm }$) and (\ref{DCijkm}), and comparing different powers of $r$ will determine the various coefficients of the expansions to any desired order. 
By doing so we arrive at
\BE
\ta{ij}{0} = \ta{ij}{1} = \ta{ij}{2} =(n-4)\ta{ij}{3}=0 , \qquad \kpi{ij}{0} = \kpi{ij}{1} = \kpi{ij}{2} = 0 , \qquad   
 \kC{ijkm}{0} =  \kC{ijkm}{1} = 0  .
\EE
In general, for $n>4$ the first non-vanishing terms are $\kC{ijkm}{2}$ and $\kpi{ij}{3}$, subject to the constraint
\BE
\kpi{ki}{3}(n-4)=\kC{ijkl}{2}b_{[lj]} . 
\label{constr}
\EE

Since we are interested here in {\em asymptotically flat} spacetimes, it is natural to demand that terms of order $1/r^2$ in~(\ref{Phi_symm}) vanish, since this is the case already in four dimensions\footnote{In the case of $n=4$ vacuum algebraically special spacetimes this is not an assumption but follows from the Ricci/Bianchi identities \cite{Sachs61}. For general vacuum spacetimes it is related to the vanishing of the unphysical Weyl tensor on scri, which holds if asymptotic flatness is assumed \cite{Penrose63}.} and the fall-off will be faster in higher dimensions. This is confirmed more rigorously (at least in even dimensions) by a study of the asymptotic behaviour of gravitational perturbations \cite{HolIsh05,HolWal04,Ishibashi08}. 
Thus from now on we set 
\BE
\kC{ijkl}{2} = 0.
\EE
By (\ref{constr}), for $n>4$ this also implies $\kpi{ki}{3}=0$. Hereafter we shall always assume $n>4$.

By comparing higher order coefficients, we obtain that for vacuum asymptotically flat $n\geq 5$-dimensional spacetimes of type II and D admitting a geodetic multiple WAND with $\det L\neq 0$, the {leading terms of the} boost weight zero components of the Weyl tensor are
\BEA
\WDS{ij} &=&\frac{\kphi{}{n-1}\delta_{ij}}{n-2}\frac{1}{r^{n-1}}+ {\cal O}(r^{-n}), \label{PhiS_leading} \qquad   
  \WDA{ij} =\frac{(n-1)\kphi{}{n-1}b_{[ij]}}{(n-2)(n-3)}\frac{1}{r^{n}}+ {\cal O}(r^{-n-1}), \ \\ 
C_{ijkm}&=&\frac{2\kphi{}{n-1}(\delta_{jk}\delta_{im}-\delta_{jm}\delta_{ik})}{(n-2)(n-3)} \label{Cijkm_leading}
\frac{1}{r^{n-1}}
+ {\cal O}(r^{-n}).
\EEA
Subleading terms can similarly be determined to any desired order \cite{OrtPraPra09inprep}, and $\kphi{}{n-1}$ and $b$ are the only integration ``constants'' characterizing the expansions~(\ref{Phi_symm}). It suffices here to observe that, by analyzing the next subleading order in~(\ref{DCijkm}) 
one finds ({if $\kphi{}{n-1}\neq 0$}) 
\BE
b_{(ij)}=\frac{b_{kk}}{n-2} \delta_{ij} . \label{sympartB}
\EE

Let us first briefly discuss the case of a non-twisting WAND (i.e., with $A_{ij}=0$), for which $b$ is symmetric.
Eqs. (\ref{sympartB}) and (\ref{inverse}) then imply
that $L_{ij}$ is proportional to the identity matrix and therefore shear-free. Hence such spacetimes belong to the $n>4$ Robinson-Trautman class and are thus of type D \cite{PodOrt06}. Furthermore, with no need of further calculations it follows from \cite{PodOrt06} that the only asymptotically flat metric within that family is the Schwarzschild-Tangherlini solution \cite{Tangherlini63}. We conclude that {\it Schwarzschild-Tangherlini black holes are the only ``asymptotically flat'' spacetimes within the class of vacuum metrics admitting a non-twisting, non-degenerate geodetic multiple WAND}.\footnote{By ``asymptotically flat'' we just mean the condition $\kC{ijkl}{2} = 0$, which is of course weaker than asymptotic flatness in the strict sense.}
This should be contrasted with the four-dimensional case, where all vacuum Robinson-Trautman solutions \cite{Stephanibook} in fact satisfy such assumptions. Note also that, although partly related, this result tells us something different from the fact \cite{Hwang98,GibIdaShi03} that the Schwarzschild-Tangherlini metric represents the unique asymptotically flat static vacuum  black hole. Finally, it is worth observing that properties of vacuum spacetimes with a non-twisting geodetic multiple WAND have been studied in \cite{PraPra08} without the assumptions $\det L\neq0$ and $\kC{ijkl}{2} = 0$.

For the twisting case, there are in general also components of the Weyl tensor with negative boost weight, which will now be analyzed. 

\subsubsection{Components of negative boost weight}

Similarly as for boost weight zero components, we expand Weyl tensor components of boost weight $-1$ as 
\BEA
 \Psi_{i} = \sum_{m=0}^p \kpsi{i}{m}  r^{-m} + {\cal O}(r^{-(p+1)}) ,  \qquad
 \Psi_{ijk} = \sum_{m=0}^p \kpsi{ijk}{m}  r^{-m} + {\cal O}(r^{-(p+1)}) . \label{Weyl-m1-exp} 
\EEA

Their asymptotic behaviour can then be determined from Bianchi equations (B.1), (B.6), and (B.9) of \cite{Pravdaetal04} 
 \BEA
D\Psi_i&=&-2\Psi_s L_{si} +\delta_i \WD{} -\WD{} L_{i1}-2\WDA{is} L_{s1} -\WD{is} L_{s1}\label{PII-B1},\\
2D\Psi_{ijk}&=& -2\Psi_{ijs}L_{sk}-\Psi_{i}L_{jk}+\Psi_{j}L_{ik}-2\delta_k \WDA{ij} +2\WDA{ij} L_{k1}-\WD{ki} L_{j1}+\WD{kj} L_{i1}+C_{ksij}L_{s1}-4\WDA{[i|s} \Ms_{j]k}\label{PII-B6},\\
D\Psi_{jki}&=&2\Psi_{[k|si}L_{s|j]}+\Psi_{i}A_{jk}-\delta_{[k}\WD{j]i} +\WD{[k|s} \Ms_{i|j]}-\WD{si} \Ms_{[jk]}.
\label{PII-B9}
\EEA
First, {it is now evident that the knowledge of the $r$-dependence of $L_{ij}$ is not sufficient in this case, and} one  needs to determine the asymptotics of the Ricci rotation coefficients $L_{i1}$ and $\Mi_{jk}$ and of  the operator $\delta_i$. This can be done using appropriate Ricci equations \cite{OrtPraPra07} and commutators \cite{Coleyetal04vsi}. These imply that, asymptotically,
\BE
L_{i1} = {\cal O}(r^{-1}), \qquad  \Mi_{jk} = {\cal O}(r^{-1}), \qquad   
\delta_i = {\cal O}(r^{0}) \frac{\partial}{\partial r} + {\cal O}(r^{-1}) \frac{\partial}{\partial x^A} , \label{delta_exp}
\EE
where the $x^A$ represent any set of ($n-1$) scalar functions (which need not be further specified for our purposes) such that $(r,x^A)$ is a well-behaved coordinate system. More details on the specific form of 
(\ref{delta_exp}) will be given elsewhere in a more general context \cite{OrtPraPra09inprep}.

Substituting  expansions (\ref{Phi_symm}) and (\ref{Weyl-m1-exp}) (with (\ref{PhiS_leading}), (\ref{Cijkm_leading}), (\ref{delta_exp})) into (\ref{PII-B1})--(\ref{PII-B9}) and comparing different powers of $r$ leads to
\BE
\kpsi{i}{0} = \kpsi{i}{1} = \kpsi{ijk}{0} = \kpsi{ijk}{1} = 0 , \qquad (n-4)\kpsi{i}{2} = 0.
\EE
The last equation  indicates  that there is a fundamental difference between four and higher dimensions. 
As a consequence, by studying higher orders in $1/r$ in the expansion of (\ref{PII-B1})--(\ref{PII-B9}), we find that 
all terms in the expansion of the Weyl tensor components up to $\kpsi{i}{n-2} $ and $\kpsi{ijk}{n-2} $ vanish for $n>4$, i.e.
\BEA
\kpsi{i}{2} = \ \dots \  = \kpsi{i}{n-2}=0 , \qquad \kpsi{ijk}{2} = \ \dots \  = \kpsi{ijk}{n-2}=0  \qquad  (n>4) . \label{van_terms_bwm1}
\EEA

Finally, the relevant Bianchi equation for determining the asymptotic behaviour of boost weight $-2$ components $\Psi_{ij}$ is equation (B.4) of \cite{Pravdaetal04}):
\BEA
2D\Psi_{ij}&=&
-2\Psi_{is}L_{sj}+\Delta \WD{ji} +\delta_{j}\Psi_i+2\Psi_i L_{[1j]}+2\Psi_{jsi}L_{s1}
+\WD{} N_{ij}\nonumber\\ && {}
-2\WDA{is} N_{sj}+\WD{si} N_{sj}+\WD{js} \Ms_{i1}+\WD{si} \Ms_{j1}+\Psi_s \Ms_{ij}.\label{PII-B4}
\EEA
We again expand $\Psi_{ij}$ as 
\BE
 \Psi_{ij} = \sum_{m=0}^p \kpsi{ij}{m}  r^{-m} + {\cal O}(r^{-(p+1)}).
\EE
Clearly, one has first to determine the asymptotic behaviour of the Ricci rotation coefficients $L_{1j}$, $\Ms_{i1}$, $N_{ij}$ and of the operator $ \Delta $. As before, appropriate Ricci identities and commutators give 
\BE
  L_{1j} = {\cal O}(r^{-1}), \qquad  \Ms_{i1} = {\cal O}(r^{0}) , \qquad  N_{ij} = {\cal O}(r^{-1}) , \qquad 
  \Delta={\cal O}(r) \frac{\partial}{\partial r} + {\cal O}(r^{0}) \frac{\partial}{\partial x^A} . \label{Delta_exp}
\EE
(Again, see \cite{OrtPraPra09inprep} for more technicalities). 
With these and (B.13) of \cite{Pravdaetal04}), one finally arrives at
\BEA
\kpsi{ij}{0} =\ \dots \  = \kpsi{ij}{n-2}=0  \qquad (n>4) . 
 \label{van_terms_bwm2}
\EEA

Note that all higher order terms of negative boost weight also vanish in the non-twisting case (which indeed is of type D, as mentioned above - cf.~also \cite{PraPra08}). In the twisting case higher order terms are possibly non-zero and can be determined to any desired order \cite{OrtPraPra09inprep} (in particular, one finds that if $\kphi{}{n-1}= 0$ the spacetime is flat). The result (\ref{van_terms_bwm1}) and (\ref{van_terms_bwm2}) is however sufficient for our analysis, as we now discuss.

\section{Discussion}
\label{sec_discussion}

To summarize, we have shown that for the considered class of spacetimes, the asymptotic behaviour of various components of the Weyl tensor along a geodetic multiple WAND is determined by eqs.~(\ref{PhiS_leading}) and (\ref{Cijkm_leading}) (which in turn can be used to recursively fix higher order terms and negative boost weight components \cite{OrtPraPra09inprep}). Perhaps surprisingly, the first non-zero components are, irrespective of their boost weight, of order $1/r^{n-1}$ (or higher). On the other hand, for radiative solutions (in even dimensions) one expects a much slower fall-off \cite{HolWal04} (cf.~also \cite{OrtPraPra09inprep}). In physical terms, our results thus imply that {\it higher-dimensional asymptotically flat vacuum spacetimes admitting a non-degenerate, geodetic multiple WAND do not contain gravitational radiation}. This conclusion is in striking contrast with the four-dimensional case, for which the Weyl tensor of a generic algebraically special vacuum spacetime whose multiple PND is non-degenerate contains a radiative contribution and exhibits the well-known peeling property~\cite{Sachs61}
\BE
\Psi_2 = {\cal O}(r^{-3}), \qquad  \Psi_3 = {\cal O}(r^{-2}), \qquad  \Psi_4 = {\cal O}(r^{-1}) \qquad (n=4) ,
\EE 
where the components of boost weight 0, $-1$, and $-2$ are represented by the standard Newman-Penrose scalars.\footnote{It is worth emphasizing again that for $n=4$ it is not necessary to assume ``asymptotic flatness'', and that any multiple PND (WAND) is necessarily geodetic in vacuum by the Goldberg-Sachs theorem.} 

We have also shown (see sec.~\ref{bw_zero}) that in the non-twisting case the multiple WAND is shear-free, which leads to the uniqueness of Schwarzschild-Tangerlini black holes within that class of spacetimes. 

The interest in the approach used in this paper goes beyond the specific results discussed above. Similar techniques can be applied also in a more general context (i.e., type I or G spacetimes) and will possibly provide some insight into properties of radiating solutions. This is now being studied and will be presented elsewhere.



\end{document}